\def\tr{\mathop{\rm tr}\nolimits}
\documentstyle[12pt,worldsci]{article}
\begin{document}

\begin{center}
{\bf The Dual Description of Long Distance QCD and the Effective Lagrangian for
Constituent Quarks}\\
\vspace{.1in}
M. Baker\footnote{Work supported in part by the U.S. Dept. of Energy under
Contract No. DE-FG06-91ER40614}\footnote{Talk given by M.~Baker at the
International Workshop on Color Confinement and Hadrons, March 22-24, 1995,
Osaka, Japan.}\\
{\it University of Washington, Seattle, WA 98105}\\
\end{center}
\vspace{.5in}
\begin{center}
{\bf ABSTRACT}
\end{center}

We describe long distance QCD by a dual theory in which the
fundamental variables are dual potentials coupled to monopole fields
and use this dual theory to
determine the effective Lagrangian for constituent quarks.
We find the color field distribution surrounding a
quark anti-quark pair to first order in their velocities.  Using these
distributions we eliminate the dual potentials and
obtain an effective interaction Lagrangian $L_I ( \vec x_1 \, , \vec x_2 \, ;
\vec v_1 \, ,
\vec v_2 )$ depending only upon the quark
and anti-quark coordinates and velocities, valid to second order in their
velocities.  We propose $ L_I $ as the Lagrangian describing the long distance
interaction of constituent quarks.

\section{\bf Introduction}

In 1976 Mandelstam and 't Hooft \cite{1} proposed that the force between a
quark-antiquark pair is analogous to the force between a monopole
anti-monopole pair in a superconductor.  A dual Meissner effect prevents the
electric color flux from spreading out as the distance between the quark
anti-quark pair increases.  As a result a linear potential develops which
confines the quarks in hadrons.  A concrete model for this ``dual
superconductor'' mechanism is realized by a dual theory \cite{2} defined by a
Lagrangian density ${\cal L}_{\hbox{\tiny eff}} (C_\mu , B_i )$
in which the
fundamental variables are dual potentials $ C_\mu $ coupled to monopole fields
denoted by $B_i$.  The local coupling of monopole fields to dual potentials
gives rise to a $C_\mu$ mass and consequently to a dual Meissner effect.
The Lagrangian density ${\cal L}_{\hbox{\tiny eff}} (C_\mu , B_i )$ of the
dual theory describes
the same long distance physics as the original QCD Lagrangian density ${\cal
L}_{\hbox{\tiny QCD}}$ defined in terms of Yang Mills potentials $A_\mu$
which are
strongly coupled at long distances.

In this talk we construct ${\cal L}_{\hbox{\tiny eff}}$, treating quarks as
classical point particles having coordinates $\vec {x}_i (t)$ and velocities
$\vec {v}_i (t)$.  We solve the resulting classical field equations
for $C_{\mu}$ in the presence of a
quark anti-quark pair to first order in their velocities.  Using these
solutions we eliminate the dual potentials $C_{\mu}$ from ${\cal
L}_{\hbox{\tiny eff}}$ and obtain an effective interaction Lagrangian $
L_I(\vec
{x}_{1},\vec{x}_{2}; \vec{v}_{1}, \vec{v}_{2}) \equiv \int d\vec{x}{\cal
L}_{\hbox{\tiny eff}}$ depending upon the quark and anti-quark
positions and
velocities, valid to second order in these velocities.  It contains two
parameters $ \alpha_s $ and the string tension $ \sigma $.
We use $ L_I $ and the canonical formalism to construct the quark
anti-quark Hamiltonian and obtain a specific quark model for calculating the
meson spectrum.~\cite{2}

We begin by showing that dual potentials give an alternate
formulation of classical electrodynamics.  This gives an example where the
dual theory and the original Maxwell theory describe the same physics at all
distances in a situation where the original Maxwell potentials and the dual
potentials can have different long distance behavior.

\section{\bf Dual Potentials In Electrodynamics}

The
inhomogeneous Maxwell equations for a pair of oppositely charged particles
moving in a relativistic dielectric medium
are:
\begin{equation}
\vec{\nabla} \cdot \vec D = \rho \, , \quad \vec{\nabla} \times \vec H =
\vec J + {\partial \vec D \over \partial t} \, ,
\end{equation}
where
\begin{equation}
\rho ( \vec x \, , t ) =  e [ \delta ( \vec x - \vec {x_1} (t)) - \delta (\vec
x - \vec x_2 (t))] \, ,
\end{equation}
\begin{equation}
\vec J (\vec x \, , t) = e [\vec v_1 (t) \delta ( \vec x - \vec x_1 (t)) -
\vec v_2 (t) \delta ( \vec x - \vec x_2(t))] \, .
\end{equation}
Introduce a line of polarization $ \vec P_s (\vec x \, ,t)$ connecting the
particles (a Dirac string):
\begin{equation}
\vec P_s (\vec x \, , t) =  e \int^{\vec x_1 (t)}_{\vec x_2 (t)} d \vec y
\delta ( \vec x - \vec y (t)) \, ,
\end{equation}
where the integral $d \vec y$ is along any path $ L (t) $ connecting $\vec x_1
(
t)$ and
$\vec x_2 (t)$. (See Fig. 1.)
\begin{figure}[h]
\vspace{2.25in}
\caption{ Dirac string connecting oppositely charged particles}
\label{fig1}
\end{figure}

\noindent
As the particles move, so does the Dirac
string.  There results a magnetization $\vec M_s (\vec x \, , t)$,
\begin{equation}
\vec M_s (\vec x \, , t) = \int^{\vec x_1  (t)}_{\vec x_2 (t)} e d \vec y
\times
{\partial \vec y \over \partial t} \delta (\vec x - \vec y (t)) \, .
\end{equation}
It is readily checked that
\begin{equation}
\rho = - \vec {\nabla} \cdot \vec P_s \, , \quad {\rm and}\quad \vec J =
\vec {\nabla} \times \vec M_s + {\partial \vec P_s \over \partial t} \, .
\end{equation}
Inserting eqs.~(6) into (1), we obtain the solution for $\vec
D$ and $\vec H$ of the form:
\begin{equation}
\vec D = - \vec {\nabla} \times \vec C - \vec P_s \quad , \quad \vec H = -
\vec
{\nabla} C_0 - {\partial \vec C \over \partial t} + \vec M_s \, .
\end{equation}
Gauss' Law and Ampere's Law (1) have thus become  kinematic equations whose
solution, eqs.~7, define dual potentials $C^\mu = ( C^0 \, , \vec C )$.

The constitutive equations $\vec B = \mu \vec H$ and $\vec E = { 1 \over
\epsilon }\vec D = \mu \vec D$ determine $\vec E $ and $\vec B$ in terms of
the dual potentials $C_\mu$ and the magnetic permeability $ \mu =
{1 \over \epsilon}$.
The homogeneous
Maxwell equations
\begin{equation}
\vec {\nabla} \cdot \vec B = 0 \, , \quad {\rm and} \quad \vec
{\nabla} \times
\vec E = -
{\partial \vec B \over \partial t} \, ,
\end{equation}
are then dynamical equations determining $C_\mu $.

As an example, consider a homogeneous medium with particles at rest.  Then the
equations $\vec {\nabla} \times \vec E = 0$, $\vec E = \mu \vec D$, and
eq.~(7) give the following equation for $\vec C$
(denoted $\vec C_{\hbox{ \tiny DIRAC}} \equiv \vec C_D$):
\begin{equation} 
\vec{\nabla}\times (-\vec{\nabla}\times \vec{C}_{D}) =
\vec{\nabla}\times \vec{P}_{s}\, .
\end{equation}
Comparing eqs.~(9) and (4) with the equation determining the vector
potential of
a
magnetic dipole we obtain:
\begin{equation} 
\vec C_D = -\int_{\vec{x}_{2}(t)}^{\vec{x}_{1}(t)}
{ed\vec{y}\over 4 \pi}\times{(\vec x - \vec y) \over {\mid\vec x - \vec y
\mid}^
3} \, .
\end{equation}
Eqs. (4), (7) and (9) then yield the electric field
\begin{equation} 
\vec{D} = -\vec{\nabla}\times\vec{C}_{D} - \vec{P}_{s} =
\vec{D}_{\hbox{\tiny COULOMB}} \equiv \vec{D}_{\hbox{\tiny C}} \, ,
\end{equation}
\noindent
where
\begin{equation} 
\vec{D}_{\hbox{\tiny C}} = \frac{e}{4\pi}\left \{\frac{\vec x -\vec
x_1}{{\mid\vec x - \vec x_1 \mid}^3} -
{\vec x - \vec x_2\over{\mid\vec x - \vec x_2 \mid}^3}\right \} \, .
\end{equation}
The first term $-\vec{\nabla}\times\vec{C}_{D}$ in eq.~(11) produces a
field
analogous to the magnetic field of a line of magnetization
flowing through the string.
The second term $-\vec{P}_{s}$ cancels
the field through the string leaving the desired pure Coulomb field.  (See
Fig.
2.)
\begin{figure}
\vspace{1.7in}
\caption{
Diagram representing string cancellation mechanism of eq.~(11)}
\label{fig2}
\end{figure}

For slowly moving charges, the equations
$\vec{\nabla}\cdot\vec{B} = 0\, , \vec B = \mu \vec H \, , \vec {\nabla}
\cdot \vec C_D = 0$ and eq.~(7) give the following equation determining
the scalar potential (denoted
$C_{0D})$:
\begin{equation} 
-\nabla^{2}C_{0D} = -\vec{\nabla}\cdot\vec{M}_{s}\, .
\end{equation}
Comparing eq.~(13) with the equation determining the scalar potential of an
electric dipole, and using eq.~(5), we obtain:
\begin{equation} 
C_{0D} = \int_{\vec{x}_{2}(t)}^{\vec{x}_{1}(t)}
{ed\vec{y}\over 4 \pi}\times\dot{\vec{y}}\cdot{(\vec x - \vec y) \over
{\mid\vec
{x}-\vec{y}\mid}^3}
\, .
\end{equation}
Eqs.~(5), (7), (10) and (14) then yield the magnetic field
\begin{equation} 
\vec H = - \vec{\nabla} C_{0D} - \frac{\partial\vec C_D}{\partial t} +
\vec M_s = \vec H_{\hbox{\tiny BIOT SAVART}} \equiv \vec H_{BS}\,,
\end{equation}
where
\begin{equation} 
\vec H_{BS} = \frac{e}{4\pi} \left[ {\vec v_1 \times (\vec x -
\vec x_1) \over {| \vec x - \vec x_1 |}^3}
- {\vec v_2 \times (\vec x - \vec x_2 ) \over | \vec x - \vec x_2|^3} \right]
\end{equation}
is the usual Biot Savart magnetic field produced by slowly moving charges.

To obtain the covariant form of the equations of the dual Maxwell theory we
define the dual polarization tensor $G^s_{\alpha \beta}$:
\begin{equation}
G_{sk}^s \equiv M_{s,k} \, , \quad \quad G_{i j }^s =
- \epsilon_{ijk} P_{s,k} \, .
\end{equation}
The field equations (8) can then be written in the form:
\begin{equation}
\partial^\alpha \mu ( \partial_\alpha C_\beta - \partial_\beta C_\alpha ) = -
\partial^\alpha \mu G_{\alpha \beta } ^s \, .
\end{equation}
These equations follow from a dual Lagrangian density
${\cal L}_C$:
\begin{equation}
{\cal L}_C = -{ \mu \over 4} ( \partial_\alpha C_\beta - \partial_\beta
C_\alpha + G_{\alpha \beta }^s )^2 \, .
\end{equation}
${\cal L}_C$ describes the same physics as the Maxwell
Lagrangian ${\cal L}_A$ :
\begin{equation}
{\cal L}_A = - {\epsilon \over 4} ( \partial_\alpha A_\beta - \partial_\beta
A_\alpha )^2 + J^\alpha A_\alpha \, ,
\end{equation}
determining the usual Maxwell equations,
\begin{equation}
\partial^\alpha \epsilon ( \partial_\alpha A_\beta - \partial_\beta A_\alpha )
= J_\beta \, .
\end{equation}
If $ \epsilon \rightarrow 0$ at long distances then
$A_\mu$ is strongly coupled (anti-shielding), but since $\mu$~$=$~${ 1 \over
\epsilon} \rightarrow \infty$, $C_\mu$ is weakly coupled
at long distances.

\section{\bf  Non-Abelian Theory}

What is the Lagrangian density
${\cal L}_{\hbox{\tiny eff}}$
describing long-distance QCD in terms of dual potentials?
Mandelstam \cite{3} has shown that we also have the freedom to use dual
potentials
$C^a_\mu$ in non-Abelian gauge theory.  't Hooft \cite{4} showed that if the
Wilson loop
satisfies an area law then the dual Wilson loop (defined in terms of dual
potentials) satisfies a perimeter law.  This is the non-Abelian analogue of
the fact that in a medium with $\epsilon < 1$, dual potentials $C^a_\mu$ are
screened at large distances.
Although in non-Abelian gauge theory there is no direct relation between
ordinary Yang Mills potentials $A^a_\mu$ and dual potentials $C^a_\mu$, we can
determine the form of ${\cal L}_{\hbox{\tiny eff}}$ by imposing the following
requirements:
\begin{enumerate}
\item ${\cal L}_{\hbox{\tiny eff}}$ must be invariant under non-Abelian
gauge transformations of the $C^a_\mu$:
\begin{equation}
C_\mu = \Omega^{-1}C_\mu \Omega + {i \over g} \Omega^{-1}\partial_\mu \Omega\,
,
\end{equation}
where $C_\mu = \sum_{a = 1}^{8} C^a_\mu {1 \over 2}\lambda_a $ \,, $ \Omega$
is
an
$SU(3)$ matrix and $g = {2\pi \over e}$ where e is the Yang Mills coupling
constant, i.e. $\alpha_s = {e^2 \over 4\pi}$.
\item ${\cal L}_{\hbox{\tiny eff}}$ generates a mass for the $C_\mu$ field
via a Higgs mechanism coupling $C_\mu$ to 3 scalar octets $B_i (i=1,2,3)$
carrying monopole charge.
\item We write a minimal ${\cal L}_{\hbox{\tiny eff}}$ satisfying (1) and
(2).
\item We assume higher dimension operators not included in ${\cal
L}_{\hbox{\tiny eff}}$ are not quantitatively important at large distances.
\end{enumerate}

First consider the coupling of a quark anti-quark pair $q \bar q$ to $C_\mu$.
We choose
a gauge where
\begin{equation}
\rho (\vec x) = e Y [ \delta (\vec x - \vec x_1) - \delta ( \vec x -
\vec x_2)] \, ,
\end{equation}
where Y is the hypercharge matrix having diagonal elements $ {1 \over 3}\, ,
{1 \over 3} \, , -{ 2 \over 3}$ and vanishing off diagonal elements
($2 \tr Y^2 = {4 \over 3}$).  Then the covariant polarization tensor
$G_{\alpha \beta}^s$ for the Dirac string connecting the quark anti-quark pair
is given by eqs.~(4), (5), and (17), multiplied by $Y$.  The dual potentials
$C_\mu$ are then also proportional to $Y$ and so $[ C_\mu , C_\nu ] = 0 $.
One unit of
gauge invariant $Z_3$ flux $e^{i \Phi}$ flows from $\bar q $ to $ q $ along the
Dirac string.  To see this, evaluate
\begin{equation}
e^{i \Phi} \equiv P e^{ i g \oint \vec C \cdot d \vec l } \, ,
\end{equation}
where we choose the path in (24) to be a small circle surrounding the Dirac
string.  Near the string the effect of the coupling to the monopole
fields $B_i$ can be neglected
and $ \vec C \sim Y \vec C_D$. Then eq. (24) becomes,
\begin{equation}
e^{i \Phi} = e^{i g e Y} = e^{i 2 \pi Y}  = e^{{2\pi i \over 3}} \, .
\end{equation}
Any continuous deformation of $ \vec C$ in $SU (3) $ leaves $\Phi$
unchanged.

We choose a gauge for the monopole fields $B_i$ so that,
\begin{equation}
B_1 = \lambda_7 B_1 (x) \, , \quad \quad B_2 = - \lambda_5 B_2 (x)
\, , \quad \quad B_3 = \lambda_2 B_3 (x) \, .
\end{equation}
As $\vec x \rightarrow \infty $, $B_i (\vec x) \rightarrow B_0 $ where $ B_0$
is
the vacuum value of the $B_i$ determined by the position of the minimum of the
Higgs Potential $W(B)$:
\begin{equation}
{ \delta W \over \delta B_i} = 0 \, ,\quad \quad{\rm at}\, B_i=B_0\, .
\end{equation}
There is no $SU(3)$ gauge transformation which leaves invariant the
vacuum values of all three $B_i$.  Dual $SU(3)$ symmetry is then
completely broken
and all the dual potentials become massive.

The explicit form of the dual Lagrangian ${\cal L}_{\hbox{\tiny eff}}$
coupling a quark anti-quark pair to $C_\mu$ is then given by,
\begin{equation}
{\cal L}_{\hbox{\tiny eff}} = 2 \tr \left\{ - {1 \over 4}
( \partial_\mu C_\nu - \partial_\nu
C_\mu + G_{\mu\nu}^s)^2  + {1 \over 2} ( {\cal D}_\mu B_i)^2 \right\} -
W \, ,
\end{equation}
where
\begin{equation}
{\cal D}_\mu B_i \equiv \partial_\mu B_i - i g [ C_\mu\, , B_i ] \, .
\end{equation}
Using eqs.~(26) and the fact that $C_\mu = Y C_\mu (x) $ we can
eliminate the color
matrices from eq.~(28) to obtain
\begin{equation}
{\cal L}_{\hbox{\tiny eff}} =  -{ 1 \over 3} (\partial_\mu C_\nu - \partial_\nu
C_\mu + G_{\mu\nu}^s)^2
+ 2 [ ( \partial_\mu B_1)^2 + ( \partial_\mu B_2 )^2 +(\partial_\mu B_3)^2  +
g^2C^2 (B_1^2 + B_2^2) ] - W \, .
\end{equation}

\section{\bf Effective Lagrangian for Constituent Quarks}

Consider a quark (anti-quark) at $\vec x_1 (t) (\vec x_2 (t) )$ moving with
velocities $\vec v_1 (t) ( \vec v_2 (t) )$ (see Fig.~3).
\begin{figure}
\vspace{1.5in}
\caption{
Quark anti-quark pair moving along trajectories $ \vec x_1 ( t )
$ and $ \vec x_2 ( t ) $, respectively.}
\label{fig3}
\end{figure}
To find the effective interaction
Lagrangian $L_I (\vec x_1 \, , \vec x_2 \, ; \vec v_1 \, , \vec v_2 )$ we
solve the classical field equations for $C_\mu$ and $B_i$ generated by ${\cal
L}_{\hbox{\tiny eff}}$, eq.~(30), to first order in $\vec v_1$ and
$\vec v_2$, and obtain the classical solutions:~\cite{2}
\begin{equation}
C_\mu = C_\mu ( \vec x \, , \vec R (t) \, , \vec v_1 (t) \, , \vec v_2 (t) )
\equiv C^{\hbox{\tiny class}}_\mu \, ,\quad \quad
B_i = B_i ( \vec x \, , \vec R (t) \, , \vec v_1 (t) \, , \vec v_2 (t))
\equiv  B^{\hbox{\tiny class}}_i \, ,
\end{equation}
where $\vec R \equiv \vec x_1 - \vec x_2 $.

For $\vec v_1 = \vec v_2 = 0$,
the scalar potential $ C^{\hbox{\tiny class}}_0 = 0 $, and
$\vec{C}^{\hbox{\tiny class}} $ and $B^{\hbox{\tiny class}}_i$
reduce to static flux tube solutions
$\vec C^{\hbox{\tiny static}} (\vec x \, , \vec R )$,
$B_i^{\hbox{\tiny static}} (\vec x \, , R )$, describing a
configuration where the color flux \cite{5} connecting the quark anti-quark
pair
is
confined to a tube of radius $ R_{FT}$ (see Fig. 4).
\begin{figure}
\vspace{1.5in}
\caption{ Static flux tube configuration}
\label{fig4}
\end{figure}
The monopole fields vanish at the center of the flux tube and approach
their vacuum value $B_0$ at large distances.  We insert this static
solution
into ${\cal L}_{\hbox{\tiny eff}}$ and integrate over all space to obtain the
central potential $V_0 (R)$:
\begin{equation}
V_0 (R) = - \int d \vec x {\cal L}_{\hbox{\tiny eff}} ( \vec C =
\vec C^{\hbox{\tiny static}}
\, , C_0 = 0 \, , B_i = B^{\hbox{\tiny static}}_i ) \, .
\end{equation}
At large $R$,  $V_0 (R) \rightarrow \sigma R$, where the calculated
string tension $ \sigma
\sim (5 B_0)^2$.
The experimental string tension $\sigma \sim .2 GeV^2$ then determines
$B^2_0$.

To first order in
$\vec v_1 (t)$ and $ \vec v_2 (t)$, the static distributions,
$\vec C^{\hbox{\tiny static}} (\vec x \, ,
\vec R )$ and $ B^{\hbox{\tiny static}}_i (\vec x \, , \vec R )$
follow the quark motion
adiabatically.  The time dependence of $\vec{C}^{\hbox{\tiny class}}$ and
$ B^{\hbox{\tiny class}}_i$ thus arises only
from the
explicit time dependence of $\vec R$.
The resulting time varying color electric
fields along with the moving Dirac string then generate a scalar
potential $ C^{\hbox{\tiny class}}_0 $ and color magnetic fields.
These produce a
velocity
dependent term $-V_2$ in the full quark anti-quark effective Lagrangian $L_I$
given by:
\begin{equation}
L_I (\vec x_1 \, , \vec x_2 \, ; \vec v_1 \, , \vec v_2 ) = \int d \vec x
{\cal L}_{\hbox{\tiny eff}} (C_\mu = C^{\hbox{\tiny class}}_\mu \, ,
B_i = B^{\hbox{\tiny class}}_i )
\, ,
\end{equation}
valid to second order in $\vec v_1$ and $\vec v_2$
since the Lagrangian is stationary about solutions to the static field
equations.

Separating $ L_I$ into its color electric and color magnetic components,
we have
\begin{equation}
-L_I (\vec R \, , \vec v_1 \, , \vec v_2 ) = V_0 (R) + V_2 ( \vec R \, ,
\vec v_1 \, , \vec v_2 ) \, ,
\end{equation}
where the static potential $V_0 (R)$ is determined by eq.~(32).
The color magnetic component $V_2$, obtained from eq.~(33), has the form
\cite{2}
\begin{eqnarray} 
V_2 ( \vec R \, , \vec v_1 \, , \vec v_2 )& = & {[\vec R \times (\vec v_1 +
\vec v_2)]^2 \over 4 R^2} V_{-} (R) +
{[\vec R \cdot (\vec v_1 + \vec v_2) ]^2
\over
4 R^2} V_{\parallel} (R)\\
& &+ {[\vec R \times {d \vec R \over d t} ] ^2 \over 4
R^2 } V_+ (R) + { (\vec R \cdot {d \vec R \over d t})^2 \over 4 R^2} V_L
(R) +
V_{\rm spin} \, . \nonumber
\end{eqnarray}

The spin dependent
potential $V_{\rm spin}$ includes both spin-spin and spin orbit
contributions.~\cite{2}  The spin orbit contribution \cite{6}
has the qualitative
features of Monte Carlo data obtained from lattice gauge theory.~\cite{7}
The potentials $V_- (R) \, , V_{\parallel} (R) \, , V_+ (R) \, ,$ and
$V_L (R)$ are moments of the field distributions determined by the
integrand
in eq.~(33).  The potentials $V_-(R)$ and $V_{\parallel}(R)$ are
related to $V_0 (R)$ by Lorentz invariance:~\cite{8}
\begin{equation} 
V_- (R) = -{1 \over 2} V_0 (R) \, , \quad V_\parallel (R) = -{ 1 \over 2 }
V_0
(R) + { R \over 2 } { d V_0 \over d R }
\, .
\end{equation}
The function $V_L(R)$ determines the energy associated with longitudinal
oscillations
$\hat R \cdot {d \vec R \over d t}$.
The function $V_+ (R)$ determines the moment of inertia $I(R)$ of the
rotating
flux tube according to the equation:
\begin{equation}
I(R) = -{1 \over2} R^2 V_+ (R) \, .
\end{equation}
At large $ R $, $ V_+ (R) \rightarrow - A R $, where $ A
\sim .21 \sigma $.
This result for $V_+ (R)$ along with the large $R$ result $V_0 (R)
\rightarrow
\sigma R$ gives a leading Regge trajectory \cite{9} $\alpha (M^2)$ which is
linear for large $M^2$ with a slope $\alpha^{'} \approx {1 \over 6.3 \sigma}
\approx .75
GeV^{-2}$.

We list below analytic parameterizations
of the integrals over the field distributions determining $V_0 (R)\, , V_+
(R)$ and $V_L (R)$:

\begin{equation}
 V_0(R) = - {4 \over 3}{\alpha_s \over R} e^{- .51 \sqrt{{\sigma \over
 \alpha_s} }R} + \sigma R - 0.65 (\alpha_s \sigma)^{1/2} \, ,
\end{equation}

\begin{equation}
   V_+ (R) = - { 2 \alpha_s \over 3 R} e^{- 1.13 \sqrt{ \sigma \over
      \alpha_s} R} - .21 \sigma R + 1.12 \sqrt{\alpha_s \sigma} \, ,
\end{equation}
\begin{equation}
      V_L(R) = -{4 \alpha_s \over 3 R } e^{-.68 \sqrt{\sigma \over \alpha_s}
      R} +
      .09 \sqrt{ \alpha_s \sigma} \, .
\end{equation}
These potentials determine generalized Wilson loops with additional
electric and magnetic field insertions and hence can be compared with Monte
Carlo calculations of these quantities once they are carried out.

For small $R$ the effective Lagrangian $L_I$ (eqs.~(34)-(40)) approaches
$L_D$, the Darwin Lagrangian (multiplied by the color factor ${4 \over 3}$)
describing the interaction of electrons and positrons after the elimination
of the
electromagnetic field \cite{10}.  Thus in the small $R$ limit $L_I$
gives the one gluon
exchange potential to order $({1 \over m_q})^2$ just as
Darwin's classical calculation gives the one photon
exchange potential to the same order.  $L_I$ can therefore be
extrapolated smoothly from the large $R$ confinement region to the
short distance perturbative
domain.  It cannot be used at shorter distances where radiative corrections
giving rise to asymptotic freedom and a running coupling constant must be
accounted for.

It should be emphasized that the potentials (38)-(40) are obtained
from the integrated
field energy associated with the color field distributions surrounding the
moving quark anti-quark pair.  As $R$ increases, the field lines are
compressed and the resulting $L_I$ decreases less rapidly then $L_D$.  At
large $R$ the color fields evolve into a flux tube distribution and give rise
to the terms in $L_I$ linear in $R$.  $L_I$ is not calculated as a
superposition of a long distance confinement contribution and a short distance
perturbative term as eqs.~(38)-(40) might suggest.  These formulae are just
analytic parameterizations of the numerical integrals described above and the
$R$ dependence of these potentials reflect the evolution of the color field
distributions with $R$.

Finally we note that the whole concept of an effective Lagrangian depending
only upon particle positions and velocities
has a meaning only when radiation can be neglected.  To estimate when
radiation is important, consider a quark anti-quark pair separated by a
distance $R$ and moving on a circular orbit with frequency $\omega$.  Then
radiation occurs if $\hbar\omega > Mc^2$, where $M$ is the mass either of the
$C_\mu$ field or of the monopole field.  (These masses are of order 500 $MeV$).
Since $\omega \sim {c \over R}$, this means $\hbar c/R > Mc^2$, or
$R < \hbar /Mc$.
Thus, for $R > {\hbar \over Mc} \sim R_{FT}$, radiation should be
suppressed and the quark
anti-quark effective Lagrangian can be used.

\section{Energy Levels of $b \bar b$ and $ c \bar c $ Systems}

Starting with $L_I$ augmented by non-relativistic quark kinetic energy terms,
we have constructed the Hamiltonian $H$ for a heavy quark anti-quark system by
the canonical procedure.  We determined the parameters $\alpha_s \, ,
\sigma \, , m_c$ and $m_b$ by finding a best fit \cite{2} to the 17
known levels of $c
\bar c$ and $b \bar b $ systems.  Our best fit parameters
are
$\alpha_s = .37$,  $\sigma = .20 GeV^2$, $m_c = 1.34 GeV$,
$m_b = 4.77 GeV$.
These values of $\alpha_s$ and $\sigma$ lead to a flux tube radius $ R_{FT}
\sim
.5 fm$.

\section{Summary}

\noindent
(i.) We have proposed that the long distance physics of QCD is
determined by a dual
theory describing the interactions of dual potentials $C_\mu$ and monopole
fields $B_i$ which are weakly coupled at long distances.  The physical
equivalence at long distances of ${\cal L}_{\hbox{\tiny eff}} (C_\mu ,
B_i)$ and ${\cal L}_{\hbox{\tiny QCD}}$ depending upon strongly interacting
$A_\mu$ allows us to use ${\cal L}_{\hbox{\tiny eff}} (C_\mu , B_i )$ to
calculate the long distance properties of QCD
(electric magnetic duality)~\cite{11}.

\noindent
(ii.) We have used ${\cal L}_{\hbox{\tiny eff}}$ to construct an effective
Lagrangian $L_I$ for constituent quarks, which leads to a definite quark model
that can both be compared with experiment and with lattice gauge theory
simulations.

\noindent {\bf Acknowledgements}

I would like to acknowledge the contributions of my collaborators James
S.~Ball and F.~Zachariasen during all stages of this work.  I would also like
to thank Professor H.~Toki and all the organizers of this workshop for their
kind hospitality.

\end{document}